\documentclass[10pt,DIV=10]{scrartcl}

\usepackage{floatrow}
\usepackage{ amssymb }
\usepackage[ruled]{algorithm2e}
\usepackage{amsmath}
\usepackage{color}
\usepackage{graphicx}
\usepackage{colortbl}
\usepackage{booktabs}
\usepackage{float}
\usepackage{xspace}
\usepackage{multirow}
\usepackage{capt-of}
\usepackage{listings}
\usepackage{amsmath}
\usepackage{url}
\usepackage{makecell}
\usepackage{textcomp}
\usepackage{siunitx}
\usepackage{caption}
\usepackage{xpatch}
\usepackage{tikz}
\usepackage[hidelinks]{hyperref}
\usepackage[capitalise]{cleveref}

\newcolumntype{L}[1]{>{\raggedright\let\newline\\\arraybackslash\hspace{0pt}}m{#1}}
\newcolumntype{C}[1]{>{\centering\let\newline\\\arraybackslash\hspace{0pt}}m{#1}}
\newcolumntype{R}[1]{>{\raggedleft\let\newline\\\arraybackslash\hspace{0pt}}m{#1}}

\long\def \@makecaption #1#2{%
  \addvspace{4pt}
  \if \@caprule
    \hrule width \hsize height .33pt
    \vspace{4pt}
  \fi
  \setbox \@tempboxa = \hbox{\@setfigurenumber{#1.}\nut #2}%
  \if \@dimgtrp{\wd\@tempboxa}{\hsize}%
    \noindent \@setfigurenumber{#1.}\nut #2\par
  \else
    \centerline{\box\@tempboxa}%
  \fi}

\lstdefinelanguage{xmosxsasm}{
  comment=[l]{\#},
  sensitive=false,
  keywords={nop, add, msync, setsr, bu, ldw, ssync, mjoin, setd, out, in, get,
  getr, ldc},
}
\DeclareCaptionFormat{listing}
  {\parbox{\dimexpr\textwidth-2\fboxsep}{\centering #1#2#3}}
\newcommand\copyrighttext{%
\footnotesize

\textcopyright\ ACM 2018. This is the author's version of the work. It is posted here for
your personal use. Not for redistribution. The definitive Version of Record was
published in \textit{ACM Transactions on Embedded Computing Systems}, Vol. 17,
No 3 (February 2018). \url{http://dx.doi.org/10.1145/3173042}.

  }
\newcommand\copyrightnotice{%
\begin{tikzpicture}[remember picture,overlay]
\node[anchor=north,yshift=-10pt] at (current page.north) {\fbox{\parbox{\dimexpr\textwidth-\fboxsep-\fboxrule\relax}{\copyrighttext}}};
\end{tikzpicture}%
}

\title{On the limitations of analysing worst-case dynamic energy of processing}

\begin{document}

\author{Jeremy Morse, Steve Kerrison and Kerstin Eder\\
University of Bristol}

\maketitle

\copyrightnotice

\begin{abstract}

This paper examines dynamic energy consumption caused by data during software
execution on deeply embedded microprocessors, which can be significant
on some devices. In worst-case energy consumption analysis, energy
models are used to find the most costly execution path. Taking each
instruction's worst case energy produces a safe but overly pessimistic
upper bound. Algorithms for safe and tight bounds would be desirable.
We show that finding exact worst-case energy is NP-hard, and that tight
bounds cannot be approximated with guaranteed safety. We conclude that
any energy model targeting tightness must either sacrifice safety or
accept overapproximation proportional to data-dependent energy.

\end{abstract}

\section{Introduction}
\label{sec:introduction}

A significant design constraint in the development of embedded systems is that
of resource consumption. Software executed on embedded hardware typically has
very limited memory and computing performance available, and yet must meet the
requirements of the system. To aid the design process, analysis tools such as
profilers or maximum-stack-depth estimators provide the developer with
information allowing them to refine their designs and satisfy constraints.

A less well studied constraint is the limited energy and power budgets that
apply to deeply embedded systems. This is a contemporary challenge for the
proliferation of such devices, particularly those that operate in isolated
environments, with limited energy availability and where the processor is the
largest consumer of energy in the system. A typical example would be a wireless
sensing device powered by battery, that has long intervals between wireless
communication, such that processor activity dominates the system's energy
consumption. The device may need to operate for a minimum period without the
battery being replaced, therefore it has a total energy budget.

Other examples are systems dependent on energy harvesting, or systems with low
thermal design points and thus have a maximum power dissipation level that may
be independent of the total energy they consume. Whether these constraints are
satisfiable can be examined with software analysis tools, and several
techniques have been developed that allow the estimation of software's energy
consumption
\cite{lopstr13,scopes15,fopara15,wcec-jayaseelan,wcec-wagemann}.

Within energy estimation, \textit{Worst Case} Energy Consumption (WCEC) has
been explored, determining the maximum amount of energy that can be consumed
during the execution of the software. In this paper, we shall study the
calculation of worst-case energy, considering only the effects that different
software and inputs can have on a system. The objective is to determine whether
it is possible to establish an upper bound on energy that is tighter than
over-estimating by, for example, using a maximum activity factor.  Such a
factor may be unachievable during the execution of a real program, because an
operand value that triggers the highest energy consumption in one instruction
may, through data dependency and other constraints, preclude subsequent
instructions from consuming their maximal energy~\cite{Hsiao1997}.

Energy is the integral of power over a given time interval. The power
dissipation of a processor can be apportioned in two parts: static and dynamic.
\textit{Static power} or leakage is the power dissipated for as long as the
component is turned on, irrespective of its internal state or any changing
inputs and outputs. \textit{Dynamic power} or switching activity refers to
power dissipation due to changes within the processor: the clock tree,
switching of gates and charging of data buses, which all consume energy.  We
express these more formally in \cref{sec:xcoresw}. Analysis of worst-case
instantaneous dynamic power has been well studied in the literature, but here
we consider worst-case energy, i.e. the integral of power over a program
execution.

Estimating worst-case energy for a particular program requires the computation
of these two distinct contributions to power dissipation. Static power is
constant in a stable operating environment (for example voltage, frequency and
temperature), therefore energy consumption due to static power is proportional
to program execution time.  Numerous techniques have been developed by the
\textit{Worst Case Execution Time} (WCET) community to address this matter
\cite{wcetsurvey}.  Dynamic power, however, has received much less attention.
Several models of how systems consume energy have characterised the dynamic
power only for specific inputs, averaged over all inputs, assumed the upper
bound of dynamic power for each
instruction~\cite{wcec-jayaseelan,wcec-wagemann} or assumed
no dynamic power at all~\cite{scopes15}.

In this paper we demonstrate that for the proportion of dynamic energy that is
due to switching caused by operand values, the calculation of the worst-case
input to a software execution is an NP-hard problem, and further, that this
quantity cannot be approximated to a useful factor.  Our proof applies to
processors in general, but in practice this proportion of data dependent energy
may be small.  We show that on an example processor, the Xcore
XS1-L~\cite{XS1-Architecture}, a cacheless deeply embedded microprocessor with
time-deterministic instruction execution, the proportion of energy that is
infeasible to analyse contributes approximately half of the processor's dynamic
power.

The rest of the paper is structured as follows: in \Cref{sec:background} we
examine the current state of energy estimation, and related work. In
\Cref{sec:xcoresw} we demonstrate the variation in dynamic power due to
switching caused by operand values, using the Xcore processor as a case study,
and consider this in the context of other embedded processors as well as more
complex devices.  \Cref{sec:formproblem} formalises the problem that we are
dealing with, which is shown to be NP-hard in \Cref{sec:frommaxsat}, and in
\Cref{sec:inapprox} we demonstrate that the problem cannot be effectively
approximated. We discuss the results in \Cref{sec:discussion}, including system
scopes for which accurate prediction of dynamic power variation due to operand
values should be considered, versus those where it is less of a concern.
Finally, we draw conclusions in \Cref{sec:conclusions}, with an outlook on
future work.

\section{Related work and background}
\label{sec:background}

This section identifies existing techniques for determining the energy
consumption of software, techniques for determining the maximum amount of
energy a program can consume, and the theoretical definition of the MAXSAT
problem for use in later proofs. We then go on to analyse the WCEC problem
in greater detail.

\subsection{Energy estimation techniques}

Given the high complexity of microprocessors, energy analysis based on hardware
designs tends to be resource intensive, and requires access to proprietary
data and tools. Research has instead focused on using empirical techniques to
model how processors consume energy. These models can then be used to estimate
the consumption of a real-world system.

One of the most popular techniques is the instruction level energy
model~\cite{tiwarimodel}. Various test patterns of instructions are executed on
a processor and their power empirically measured, leading to a model of per
instruction energy costs and the dynamic cost of switching between different
instructions. Simulating an instruction sequence, or interpreting a trace of an
execution, can then be combined with this energy model to produce a cost value
for the execution. \cite{Steinke2001} extend this model to include the costs
of circuit switching in instruction operands. These costs include the amount of
switching occurring on data buses supplying input operands to an instruction,
and the switching on the output when a result is written back to the register
file.

Further modelling techniques for dynamic power go beyond the core part of the
processor, such as analysing flash memory~\cite{Pallister2014},
caches~\cite{Chandra2008} and DRAMs~\cite{Lee2011}. High performance processors
feature hardware-provided counters that record metrics such as cache hit rates,
which can be used by appropriately parametrised energy
models~\cite{highlevelmodels}. The energy consumption on the buses to these
components can also be influenced by data values and can be modelled
accordingly.

In this paper, however, where deeply embedded devices without such features are
the focus, we choose to only examine the dynamic power attributable to the core
part of the processor. In larger, more complex processors, the dynamic power
due to operand values that we explore remains valid, but is a smaller
proportion of the total power dissipation, and therefore may be a lower
priority for analysis or optimisation.

\subsection{Worst Case Energy Consumption (WCEC) analysis}
\label{sec:wcec}

WCEC is a form of energy estimate, where the aim is to find the maximum amount
of energy that a piece of software will consume, without needing to execute
that software.  The problem is thus made of two parts: modelling the energy
consumption of the software under test, and searching for the execution of it
that will lead to the greatest amount of energy consumed. This problem is
defined in a similar way to the Worst Case Execution Time (WCET)
problem~\cite{wcetsurvey} where the execution time of software is modelled, and
then the longest possible path found. However, the techniques required to
obtain a solution have a number of differences.

The first publication to provide a technique for computing the WCEC of software
was by \cite{wcec-jayaseelan}, where upper bounds on the energy consumption of
several programs were inferred using energy models of software basic blocks and
an ILP solver to find a maximal path through the program. The authors
additionally debunk the suggestion that the execution path consuming the most
time is always the path that also consumes the most energy. With regards to
dynamic power, the authors assume maximal circuit switching on every clock
cycle but model power management techniques within the processor such as clock
gating to create a realistic energy model. The dynamic power of switching due to
operand values is not specifically considered, and indeed the authors
show that that its contribution of dynamic power to overall energy is low,
thus their approximation does not introduce significant imprecision. We address
the contribution of operand values to dynamic power in \Cref{sec:xcoresw}.

Resource analysis techniques that extract cost relations from programs have
been employed to analyse energy consumption bounds~\cite{lopstr13,scopes15}.
The costs used in these analyses represent energy consumption and are based on
models that provide a single energy cost per instruction, obtained by averaging
the energy measured from processing random data, constrained to yield valid operands for the respective instruction~\cite{steve-tecs}.
However, bounds obtained in this way cannot be considered safe, as executions
would exist where the energy from operand values exceeds the average case.

More recently, \cite{wcec-wagemann} have presented techniques for estimating
over and under approximations of WCEC through implicit path enumeration and
genetic algorithms, respectively. They do not, however, comment on dynamic
power at all: their absolute instruction energy model appears to assume maximum
switching for each instruction cost. Their relative energy model does not
consider real energy costs, instead estimating the difference in energy
consumption between instructions, again with no explicit consideration of dynamic
power.

Both Jayaseelan and W\"agemann identify inefficiency as being a reason why they
cannot compute accurate switching activities for circuits. As we will show in
this paper, the problem is infeasibly complex under the P $\ne$ NP assumption.

\subsection{Existing complexity results}

Switching activity is a matter studied in detail by the VLSI community for
circuit design, as the maximum instantaneous switching in a circuit can affect
the power supply requirements~\cite{realisticpower}. This problem has been
shown to be NP-hard~\cite{onthecomplexity} and numerous techniques have been
developed to estimate of the worst-case power consumption~\cite{peakpower},
allowing maximum power analysis.

Power estimation itself does not directly correspond with energy estimation.
The objective of WCEC is finding the maximal amount of circuit activity over a
time interval, rather than the instantaneous maximum, which itself may be
incompatible with the circumstances that lead to maximum energy. In particular,
software requires that computations be consistent with past inputs, creating
additional constraints and dependencies.

Switching between instructions is a notable contributor to energy consumption,
which can be controlled through the order in which instructions are executed.
Techniques have been developed to reduce consumption through instruction
scheduling~\cite{Parikh2004}, but this is known to be an NP-hard problem.
Instruction scheduling uses pre-computed costs of switching between
instructions to determine an optimal static schedule. It does not consider the
operands to instructions or any cost that does not have a fixed value.

None of these complexity results are directly applicable to the estimation of
energy in data-dependent switching during software execution. To the best of
our knowledge, we believe this is the first work to consider data-dependent
switching costs.

\subsection{Maximum satisfiability}

Part of our proof in this paper relies on demonstrating that an NP-hard problem
can be represented within the problem of calculating the amount of circuit
switching in a program. We therefore assume the reader is familiar with the
SAT problem (a full treatment of which can be found in \cite{satbook2013}),
and remember the definition of the MAXSAT form of the problem, which we embed
within operand value switching in \Cref{sec:frommaxsat}.

Briefly, the \textit{Maximum satisfiability} problem
``MAXSAT''~\cite[pp.613--631]{sathandbook} takes a set of Boolean variables,
constraints on their values in the form of a set of clauses, and finds the
variable assignment that makes the maximum number of clauses true. Unlike SAT,
not all clauses need to be true. MAXSAT is known to be NP-hard.

Formally, following the presentation of \cite{Johnson}, define $L$ to be a
set of literals, and $C$ a set of disjunctive form clauses:
\begin{align}
  \nonumber L &= \bigcup_{i>0}{\{x_i, \overline{x}_i\}} \\
  \nonumber c \in C, c &= \{l_1 \vee ...\vee  l_n \ |\  l_i \in L\} ,
\end{align}
where each $x_i$ is a Boolean variable. A truth assignment defines each $x_i$
or its negation to be true. A clause is deemed to be satisfied if at least
one literal in the clause is assigned true. A MAXSAT problem is a set of
literals and set of clauses $\langle L, C \rangle$, such that the solution is the truth assignment
that causes the maximal number of clauses to be satisfied.

\subsection{WCEC background}
\label{sec:wcecbackground}

The worst-case {\em energy} consumption problem goes beyond the worst-case
execution time problem, because the execution time of a single instruction is
largely independent of its input data. This is because timing variability has
mostly been eliminated ``by design'' through the use of synchronous logic and
the limited propagation time associated with executing individual instructions.

Despite this, there are scenarios where WCET is subject to timing
anomalies~\cite{Lundqvist:1999:TAD:827271.829103}. These arise when a seemingly
shorter execution path later results in an overall increase in time. A cache
miss may cause the worst execution time locally, but this state may later
preclude a subsequent scenario that, with a global view, would in fact be the
worst-case.

A comparable anomaly in energy is two equal-time paths, where one path contains
an instruction known to produce the worst-case energy. However, the input data
that reaches that instruction does not achieve that worst-case, due to
transformations performed upon it by previous instructions in the path.
Instead, the sum of instruction energies on the alternate path is higher than
the sum of energies on the apparent worst-case path.

In real-time embedded systems, timing-predictable processors execute
instructions within a fixed number of clock cycles, irrespective of the data
the operation works on. This is particularly beneficial to WCET analysis, which
can then focus on identifying the worst-case execution path which is determined
by the control flow, rather than by the data flow of the computation.
More advanced micro-architectural features, such as early-out of operations, or
cache hierarchies, provide higher average performance at the cost of
predictability.
This makes WCET analysis far more challenging, as tight bounds firmly rely on
timing predictability of the target architecture~\cite{thiele:2004}. 
However, even operations that have a variable execution time, such as
serialized integer multiply and divide, or floating point operations, can be
quantized by the processor's clock period into a tractable number of discrete
possibilities. The range may be in the order of tens, hundreds, or thousands of
cycles, depending on the type of operation. This extends into other
architectural features, such as caches and branch predictors, which although
more complex to analyse, can still be quantized.

Energy depends on both the execution time and the power dissipation of the
operation.
Power is not quantized in terms of the clock period, but could be considered in
terms of the number of transistor and interconnect state changes (i.e.\
switches) that \emph{may} take place during an operation, depending on the data
to be processed. The number of possible power dissipation levels is thus the
size of the powerset of the number of transistors in the device.
This is several orders of
magnitude larger than the number of timing possibilities that need to be
explored by WCET analysis. This view is itself simplified, as it does not
consider the continuous variations in temperature and voltage that a device
will be affected by.

For the techniques that are used in WCET to be directly transferable to WCEC, a
set amount of energy per operation would need to be specified and realised in
hardware, similar to specifying and ensuring, through timing analysis at design
time, that each operation fits into a fixed number of clock cycles. 
Consider the converse: A processor that presents a similar WCET analysis
difficulty would be an asynchronous design, where the precise execution time is
a non-trivial function of an operation's input data.  Such devices may have an
average delay, but actual performance or tight bounds for a given use case may
be harder to determine~\cite{async-timing}.

Energy estimates can also have varying degrees of accuracy. The strongest is to
determine the \textit{exact} WCEC, which consists of a program trace causing
the maximum amount of energy that can expended, and the input that causes the
trace. The computational difficulty of finding the exact worst-case has fuelled
interest in approximation algorithms \cite{approxalgs} that guarantee to find
an example of the energy consumed by the program that is within some factor of
the worst-case. This estimate can be considered to be an approximation with a
limited amount of uncertainty, or alternately the approximate energy plus the
uncertainty factor can be used as a safe upper bound on the true WCEC.  If the
uncertainty factor is low, this may be sufficient to prove that a design
constraint is met.

Finally, upper bounds can be derived from the structure of the program itself
through very coarse over-approximation, for example by assuming a program
always exhibits the maximum activity factor. Such over-approximation is the
most inaccurate of such techniques, but also the most feasibly achievable.
The distinction between approximation algorithms and coarse over-approximation
is that the former has a definite relationship with the exact WCEC, while the
latter is based for example only on the length of the program, and provides no
guarantees on what the exact WCEC may be.

This paper addresses the first two forms of WCEC estimate and demonstrates that
they are not feasibly computable, leaving only the coarse over-approximation
as a viable technique in practice.

\section{Circuit switching on Xcore}
\label{sec:xcoresw}

Prior WCEC papers have relied on the suggestion that the variation in dynamic
power is small in relation to other energy costs in a processor, at
approximately \SI{3}{\percent}~\cite{lowcircuitcost}, therefore a conservative
average-based model may be suitable. Other work has presented a mixed picture:
\cite{energyawaresoftware} found that the switched capacitance (i.e.  switching
cost) of a StrongARM processor had little variance across applications,
suggesting that switching costs contribute little to overall program energy;
while \cite{highdatadep} observe that data switching accounts for up to 50\% of
processor core energy.

Here, we affirm two properties of dynamic power dissipation in a processor by
analysing the Xcore~\cite{XS1-Architecture} XS1-L. First, that dynamic power
due to operand values can be high, and second, that this cost can vary
significantly.

\subsection{Defining power dissipation in a micro-processor}

The energy, $E$, of an electronic device is the integral of its power dissipation, $P$, over a given time period, $T$:
\begin{equation}
  E = \int_{t=0}^{T} P(t)\ dt \text{.}
  \label{eq:energy_integral}
\end{equation}

Power is an instantaneous measure of the rate of work. Typically, this is
sampled repeatedly in order to discretise the integral, or the power is
averaged, simplifying the equation to $E = P \times T$. In digital devices such
as processors, the total power dissipation of the device, $P_{tot}$ is
typically apportioned into two additive parts, termed static and dynamic,
denoted here as $P_s$ and $P_d$ respectively:
\begin{equation}
  P_{tot} = P_s + P_d .
  \label{eq:totalpower}
\end{equation}

Elaborating on these, static power is determined by the operating voltage,
$V_{dd}$ of the device and $I_{leak}$, the leakage current present, which is
itself dependent upon physical characteristics such as operating temperature,
transistor feature size and the manufacturing process that is
used~\cite{butzen2006leakage}, yielding an exponential
equation~\cite{gonzalez1997supply}: 
\begin{equation}
	I_{leak} = WI_se^{\frac{V_{th}}{V_{dd}}},
\end{equation}
for transistor width, $W$, sub-threshold current, $I_s$, as well as the voltage
threshold and operating voltage, $V_{th}$ and $V_{dd}$, respectively. However, a
linear approximation is sufficient for the normal operating range of many
devices, therefore a simplified representation of static power is:
\begin{equation}
  P_s = V_{dd}I_{leak},\quad \therefore \quad P_s \propto V_{dd}
  \label{eq:staticpower}
\end{equation}

Dynamic power is dependent upon the capacitance of the components that are
being switched, $C_{sw}$, as well as the operating voltage and the frequency of
switching, $f$. In a processor, $f$ is governed by the clock frequency. The
proportion of the device that is switching is dependent upon the instruction
and data being executed and related changes in state. This is represented by an
\emph{activity factor}, $\alpha$, where each instruction or action performed by
the processor may have a different $\alpha$.
\begin{equation}
  P_d = \alpha C_{sw} V_{dd}^2 f,\quad \therefore \quad P_d \propto V_{dd}^2
  \label{eq:dynamicpower}
\end{equation}

There is a quadratic relationship between voltage and dynamic power. The
necessary operating voltage is approximately linearly proportional to the
operating frequency in processors operating above the threshold voltage of
their transistors~\cite{kim2003leakage}.

\subsection{Apportioning dynamic power}

The power dissipation of a single instruction is typically expressed as the
average power observed during the execution of that instruction. When
considering a model of power per instruction, it is important to calculate an
appropriate $\alpha$ per instruction, or some equivalent by abstraction.
However, the instruction is not the sole influence upon the $\alpha$ value. The
operands supplied to the processor's functional units (for example, arithmetic
unit), will affect the amount of switching. This includes changes to the input
and output of the functional unit, as well as internal switching within the
unit as the new result is computed. As such, one instruction may have a range
of possible $\alpha$ values that are dependent on the input data.

Prior work~\cite{wcec-jayaseelan} has suggested that this variation in $\alpha$ is small
and therefore not significant enough to consider when constructing a worst-case
energy model.
However, we demonstrate that variation in input data can be responsible for
as much as \SI{42}{\percent} of a core's power dissipation
and thus becomes a
relevant contributor to the model. This is pertinent to systems with minimal
additional components, such as those that are deeply embedded, where the
processor is the major consumer of energy. In larger, more complex systems,
with multi-layer memory hierarchies, many external peripheral devices and
several power supplies, the variation in total system energy due to operand
values is proportionally smaller.

Internal processor data buses are one of the largest contributors to dynamic
power in an embedded processor. These buses interconnect various internal
units, and so changing values on these buses indicate the charge or discharge
of connections between a number of gate inputs and outputs, which may have
different loads depending on their fan-in or fan-out and connection length. The
\cite{Steinke2001} energy model explores this and discovers that approximately
\SI{20}{\percent} of overall processor power can be attributed to the Hamming
distance between transitions on buses.

To determine the dynamic power cost on our target device (the Xcore XS1-L), we
performed experiments in the manner of \cite{Steinke2001}. For a set of
instructions, we tested every combination of input operand values from zero to
255 for each operand, creating a sequence of tests, $\mathbb{P}$. We alternate
between instructions with this data set and all-zero operands, to ensure we
measure the Hamming weight on each cycle. On the Xcore, this instruction
sequence is achieved with four parallel, tightly coupled threads that are
issued into the pipeline round-robin by the processor's deterministic hardware
scheduling. Alternating threads perform the desired operation on all-zero data
and the operand values under test. Unrolled sequences are used to minimize
noise from loop instructions that we do not want to measure. In a
single-threaded in-order processor, a similar sequence can be achieved with one
thread.

An example instruction trace is given in~\cref{lst:instrseq}. This method
effects a change in energy consumption due to operand values along the core's
internal datapath, starting at the reading of the register file, progressing
through the functional units of the core, then once again to the register file
for write-back. The effect of memory access does not fall within the scope of
this work, although we acknowledge the impact of this in
\cref{sec:wcecbackground} and discuss the proportional contributions at the
system level in~\cref{sec:discussion}.

\begin{lstlisting}[label=lst:instrseq,caption={Example trace of instructions.},float]
# r0 initialised to zero, r1 and r2 initialised to desired operand
# values for each test.
# OP is desired operation, e.g. add, sub, ...
# First operand is destination register (e.g OP dest, src1, src2)
OP r0, r0, r0
OP r3, r1, r2
OP r0, r0, r0
OP r3, r1, r2
  ...         # Repeat until sufficient power samples are collected
\end{lstlisting}

The Xcore is a cache-less multi-threaded processor with time-deterministic
execution, therefore there is no complex memory hierarchy to influence power or
time, and all of the instructions that we test take the same, constant
execution time. Test sequences were constructed in such a way to ensure we
exercised the datapath between the register file and functional units in every
instruction cycle. Tests are applied in a single-core context, with no
inter-thread communication. Although the processor has 32-bit operands,
exhaustive testing over 8-bit data presents data-sensitive energy patterns that
are sufficient to expose the behaviours of interest to this work. For wider
data, these patterns repeat, potentially with larger patterns becoming
apparent. This is also evident for narrower data, where patterns are still
observable in smaller subdivisions of the figures that we present in the
following subsection. As such, using the full width of the processor is not
necessary to analyse the problem that we present in this paper.

The device is operated with a \SI{1.0}{\volt} core power supply and
\SI{500}{\mega\hertz} clock frequency. Power is sampled at the \SI{3.3}{\volt}
input to the DC-DC converter that supplies the cores and is done using a
vendor-supplied sampling and debug device that uses a shunt resistor to
determine current. The tests are each run repeatedly for a \SI{0.5}{\second}
duration in order to acquire several thousand power samples, then taking the
average.

The device under test is the XS1-A16A-128, a dual-core component, tested with
single-core code. The device has two discrete cores in a single processor
package, with voltage regulators and unused analogue peripherals being the only
shared components also measured. We assume the peripheral components consume
negligible energy when unused, and that the power supply contribution will be
proportionally similar on single core, assuming the DC-DC is re-selected to
achieve the same efficiency at a lower current.

We remove the additional energy consumption that would not be present if a
single-core version of the component were to be used. We refine the total
power, $P_{tot}$, by considering the distribution of power between the two
cores.  This is established through a simple step, measuring the total power of
the dual-core device when idle, represented as $P_{tdual}$, then enabling us to
ascertain $P_{tsingle}$, representing a single core:
\begin{equation}
  P_{tsingle} = \frac{P_{tdual}}{2} .
\end{equation}

Following this, the dynamic power dissipation due to operand values is
isolated. Executing instruction power tests on one core, leaving the remaining
core idle, produces a sequence of test results, $\mathbb{P} = \{P_0, \dots, P_n
\}$. We define the dynamic power contribution of the lowest and highest power
test cases as $P_{dmin}$ and $P_{dmax}$ respectively, and the dynamic power
range, $P_{drng}$:
\begin{align}
  P_{dmin} &= \min(\mathbb{P}) - P_{tdual},  \\
  P_{dmax} &= \max(\mathbb{P}) - P_{tdual},  \\
  P_{drng} &= P_{dmax} - P_{dmin},
\end{align}
or as a percentage contribution to total processor power:
\begin{equation}
  \frac{P_{x}}{P_{tsingle} + P_{x}} \enskip , \enskip \forall x \in \{dmin, dmax\} .
  \label{eq:pctrange}
\end{equation}

We observe for the device under test that $P_{tdual} = \SI{328}{\milli\watt}$
and therefore $P_{tsingle} = \SI{164}{\milli\watt}$. Any additional power
observed during tests is used to determine how much dynamic power variation is
possible for the set of input values tested. This is not solely static power,
because even at idle, switching in components such as the clock tree is taking
place, contributing to dynamic power. Thus, the difference in power observed
during instruction and data tests is not the total dynamic power contribution,
but does establish the degree of variation in dynamic power that can take
place, and what proportion of total core power this amounts to.

\subsubsection{Heat-map observations}

To aid analysis of the results of these experiments, we present a series of
``heat-map'' figures, showing measured dynamic power in colour, and with
operand (input) and result (output) Hamming weights in greyscale. These plots
use measurements from tests of the \texttt{add} instruction, although similar
data-dependent behaviours were observed in other instructions.

\begin{figure}
  \begin{floatrow}
    \ffigbox[\FBwidth]{\includegraphics[width=0.47\textwidth,clip,trim=0.75cm 0cm 2.4cm 1cm]{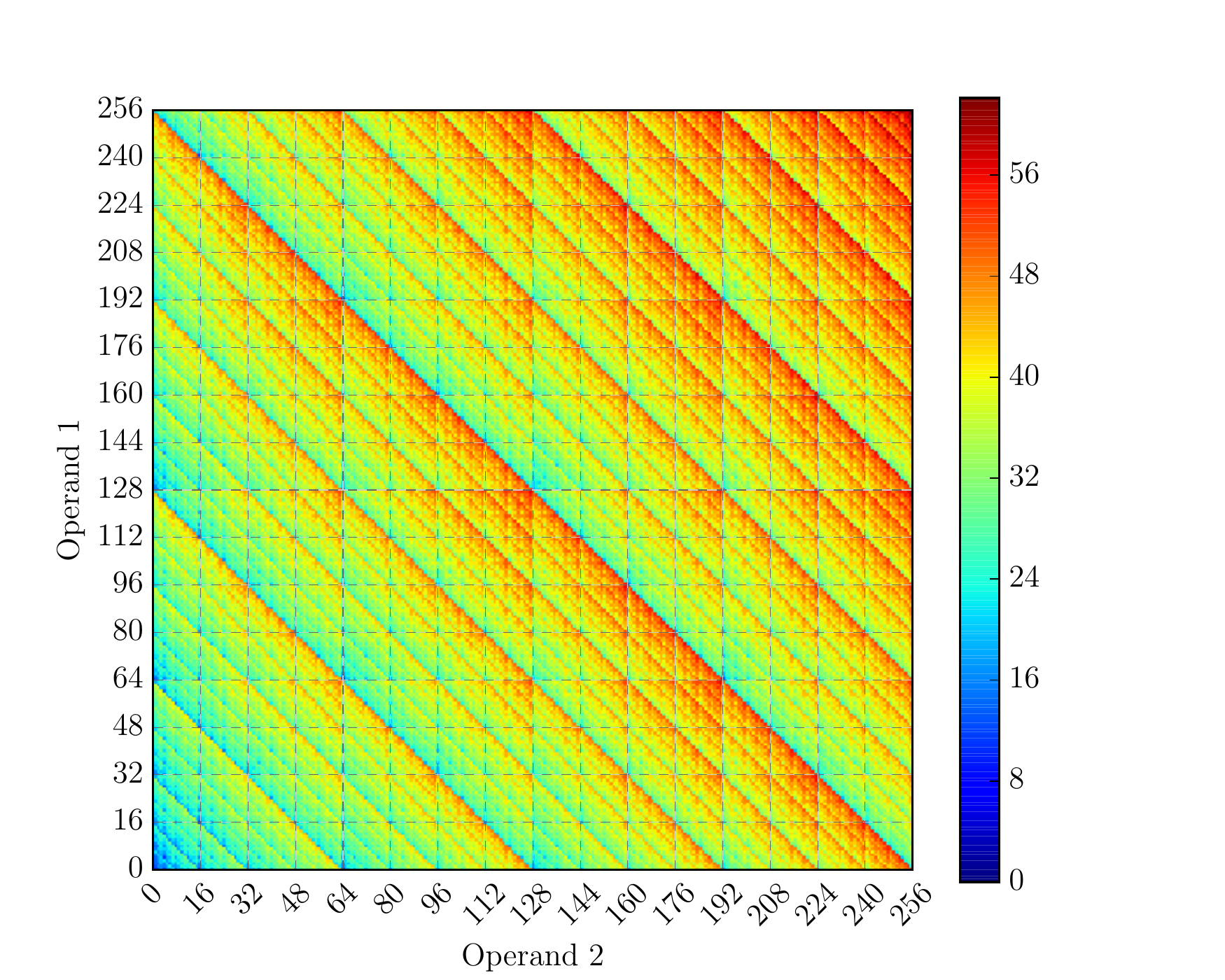}}{\caption{Dynamic power in milliwatts for \texttt{add}, over a range of input operand values.}\label{fig:heatmap1}}
    \ffigbox[\FBwidth]{\includegraphics[width=0.47\textwidth,clip,trim=0.75cm
    0cm 2.4cm 1cm]{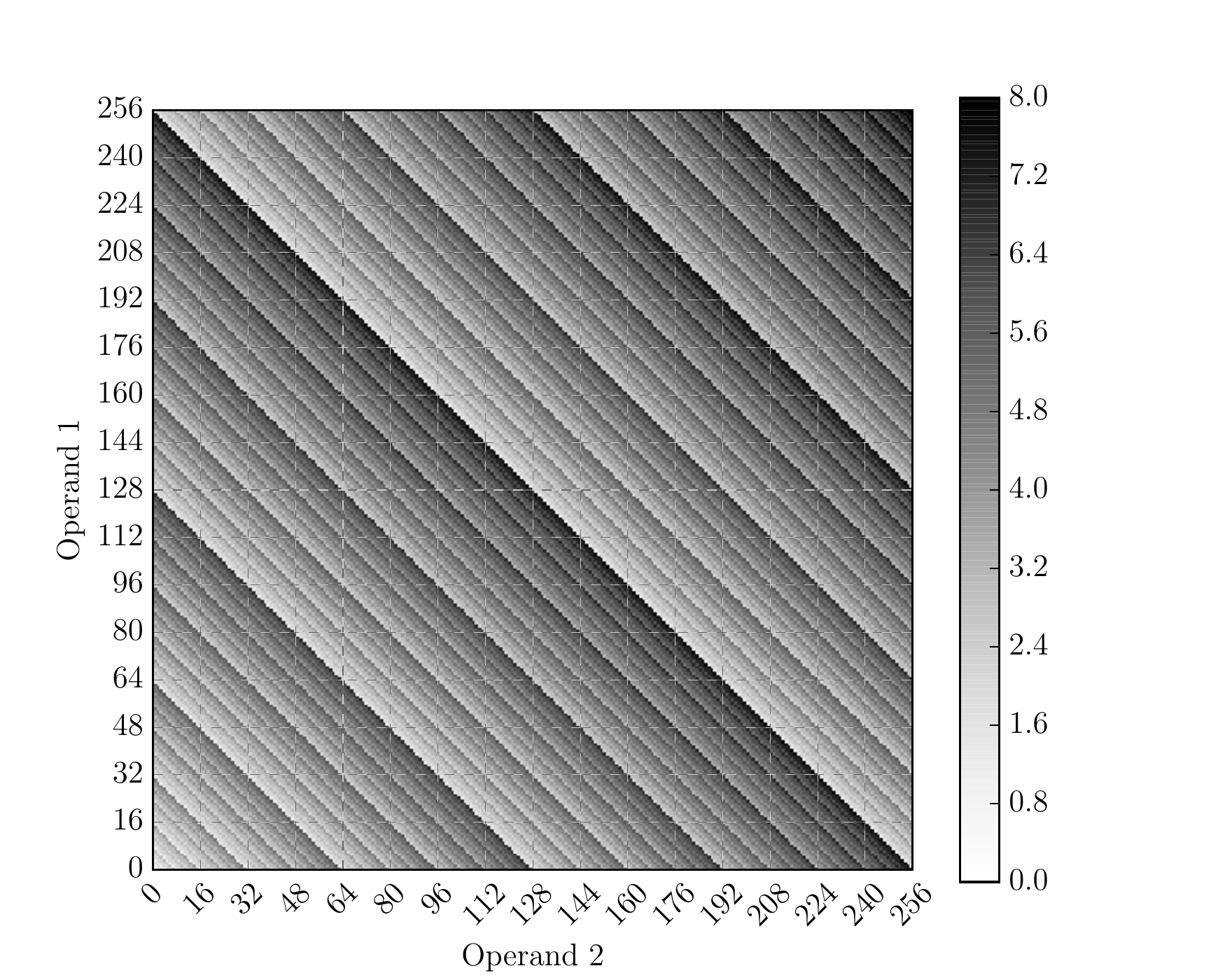}}{\caption{Hamming weight of the output value from
    performing \texttt{add}, in number of bits set.}
    \label{fig:sumhammingweight}}
  \end{floatrow}
\end{figure}

\Cref{fig:heatmap1} shows total dynamic power for \texttt{add} with all
combinations of two 8-bit operands. The diagonal striping indicates a strong
correlation with the number of bits set to \texttt{1} in the result of the
computation.  This is observable due to alternating between test \texttt{add}
operations and operations with all-zero inputs and outputs. The Hamming weight
of the result values are shown in \Cref{fig:sumhammingweight}. This is
determined to represent \SI{4.4}{\milli\watt} per output bit set in respect to
the data from \Cref{fig:heatmap1}.
Using the Hamming weight range as a scale from zero mW to the maximum mW
observed, the Hamming weight predicts dynamic power with a mean error of 3mW.

\begin{figure}
  \begin{floatrow}
    \ffigbox[\FBwidth]{\includegraphics[width=0.47\textwidth,clip,trim=0.75cm 0cm 2.4cm 1cm]{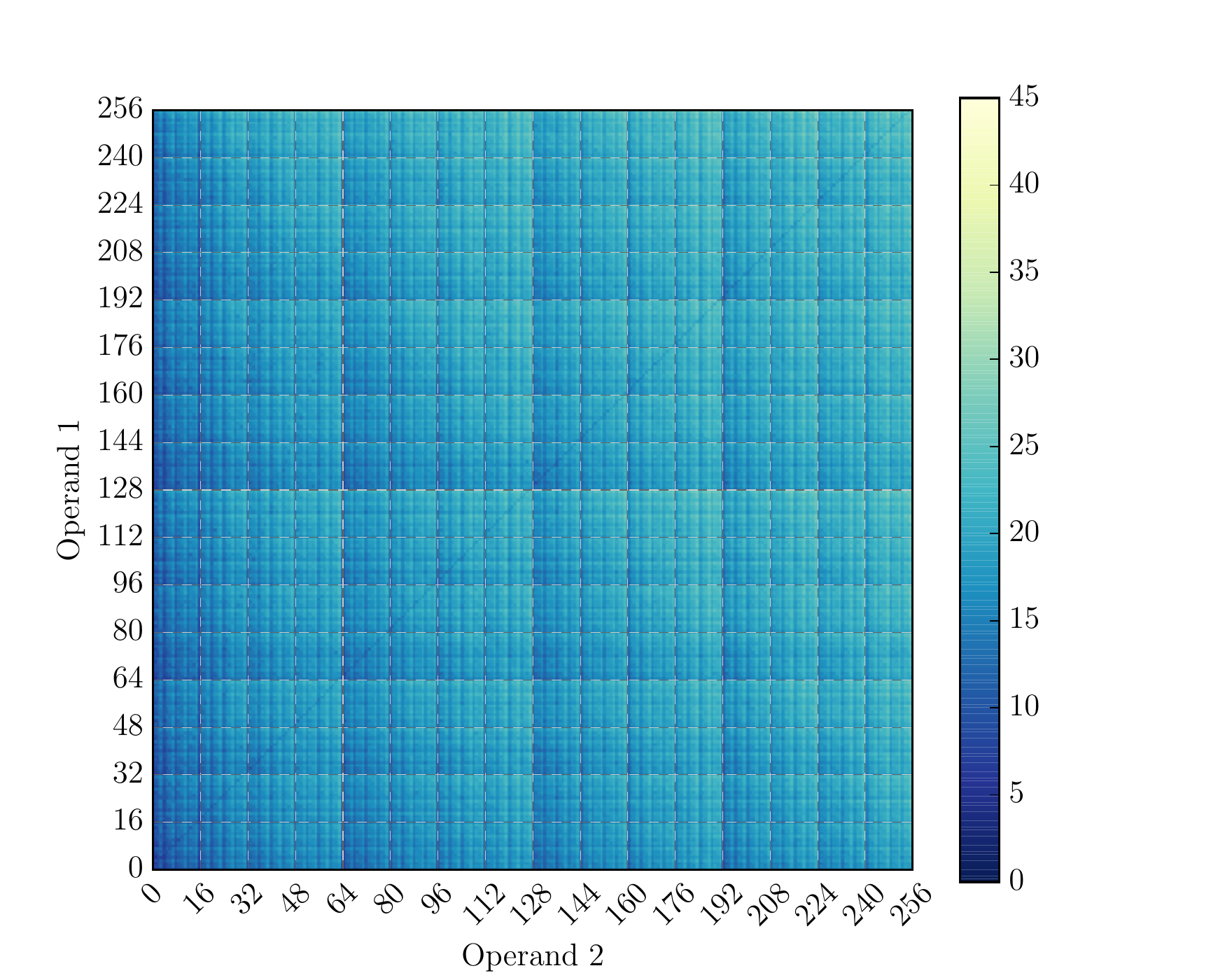}}{\caption{Dynamic power in milliwatts for \texttt{add}, with the output value cost subtracted, at 4.4mW per bit.}
    \label{fig:heatmap2}}
    \ffigbox[\FBwidth]{\includegraphics[width=0.47\textwidth,clip,trim=0.75cm 0cm 2.4cm 1cm]{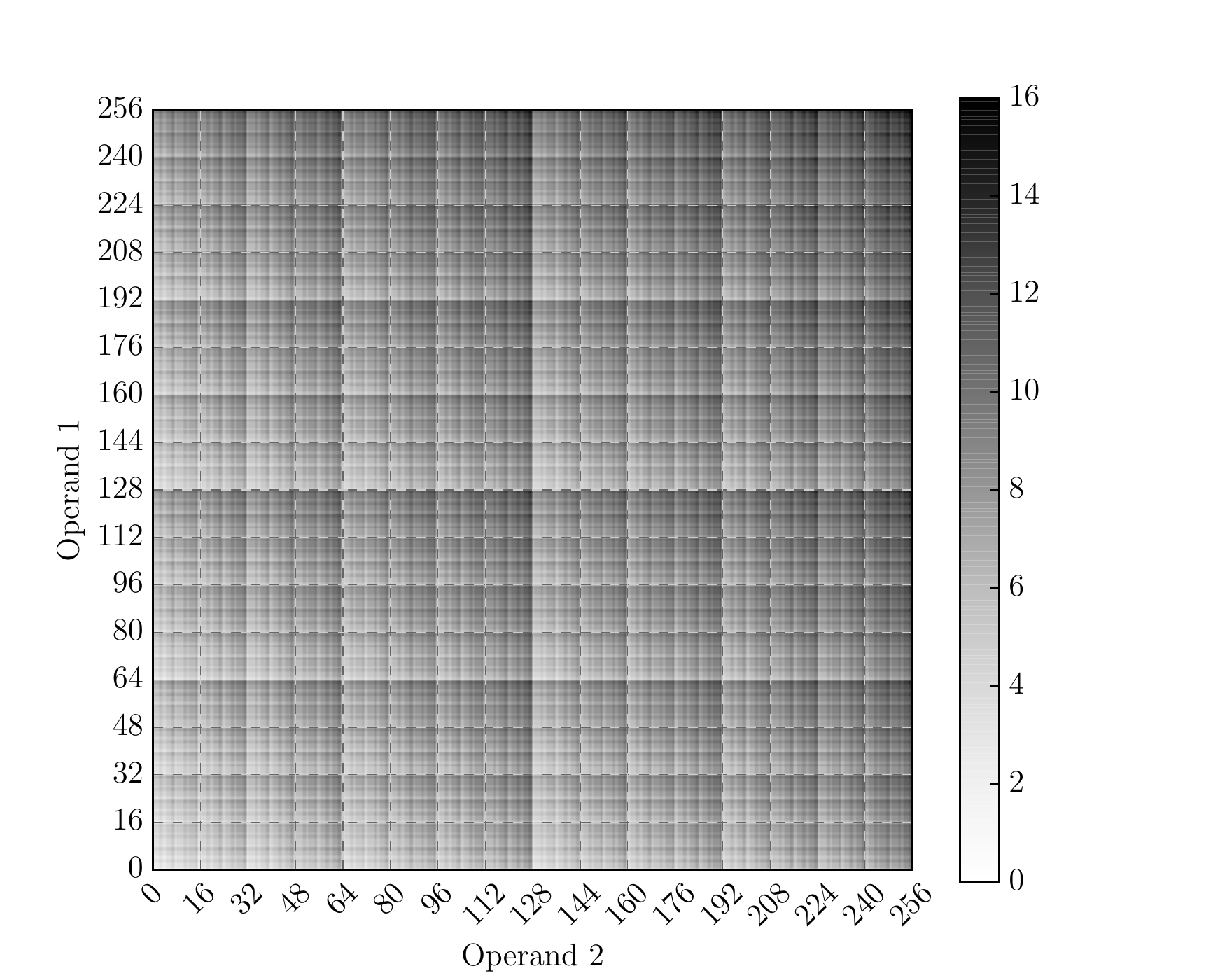}}{\caption{Hamming weight of both input operands to \texttt{add}, in number of bits set.}
      \label{fig:inputhammingweight}}
  \end{floatrow}
\end{figure}

\begin{figure}
  \begin{floatrow}
    \ffigbox[\FBwidth]{\includegraphics[width=0.47\textwidth,clip,trim=0.75cm 0cm 2.4cm 1cm]{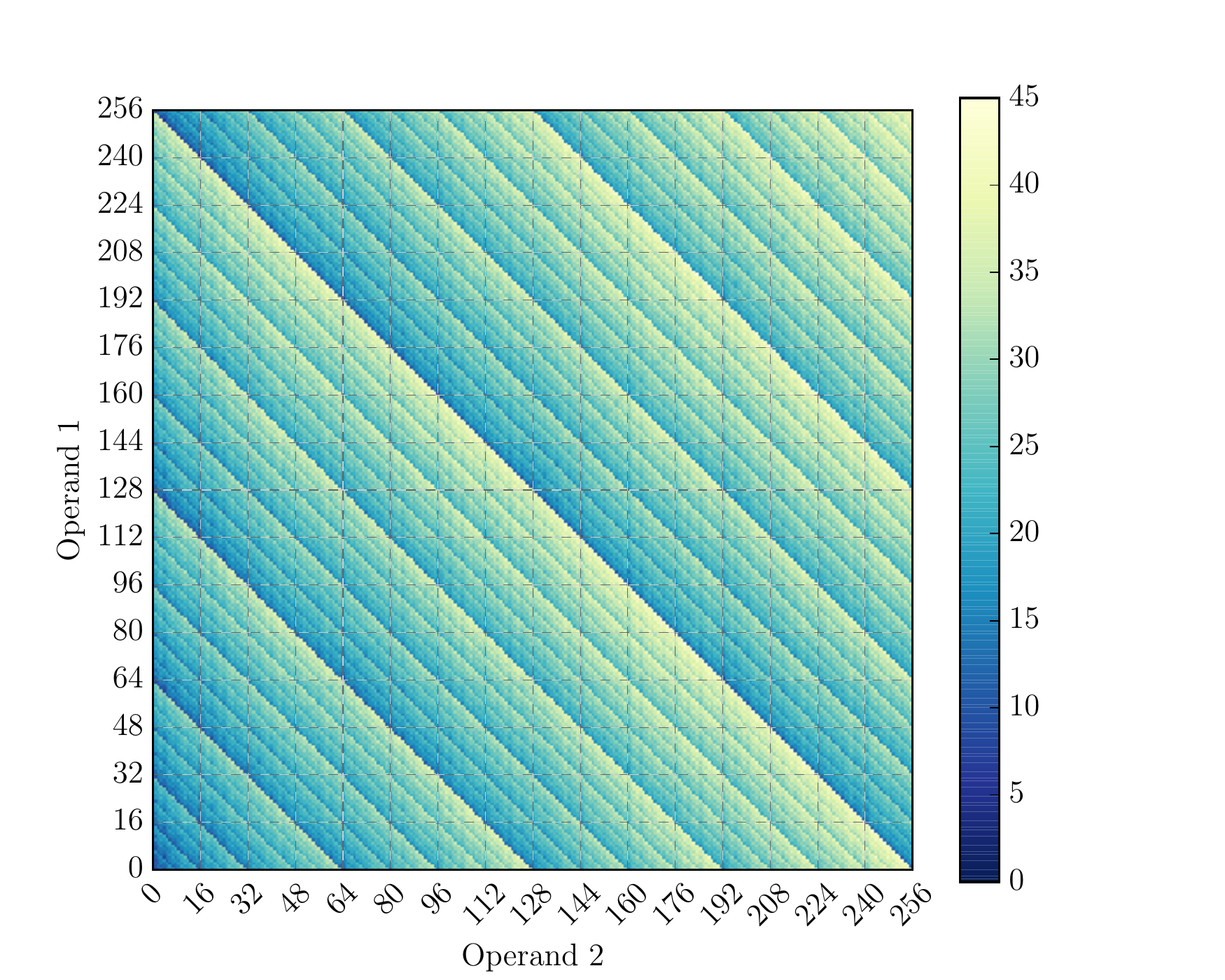}}{\caption{Dynamic power in milliwatts for add instruction, with input datapath cost subtracted (assuming 1.3mW per bit).}
    \label{fig:heatmap3}}
    \ffigbox[\FBwidth]{\includegraphics[width=0.47\textwidth,clip,trim=0.75cm 0cm 2.4cm 1cm]{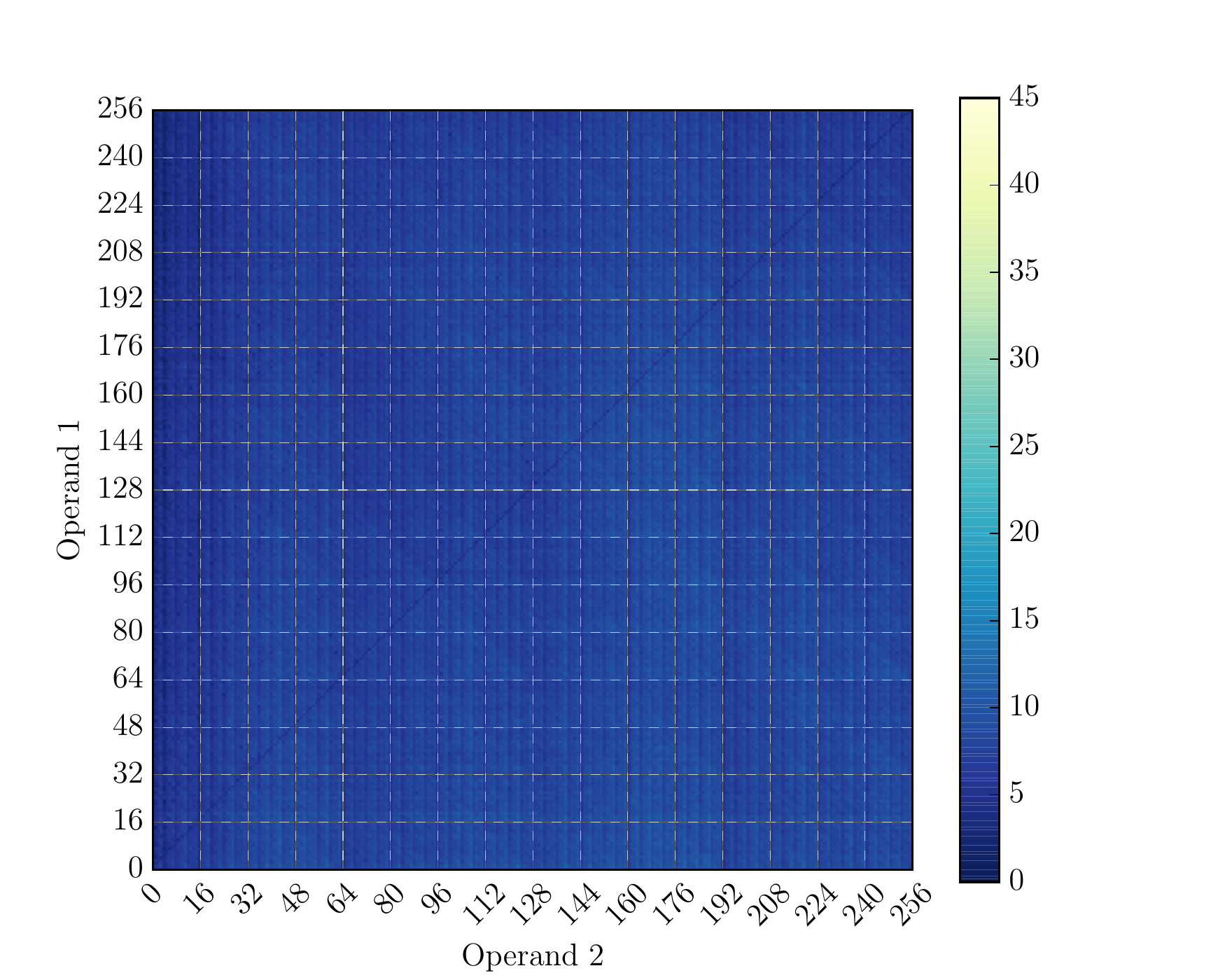}}{\caption{Dynamic power in milliwatts for add instruction, with operand and result value costs subtracted.}\label{fig:heatmap4}}
  \end{floatrow}
\end{figure}

Subtracting the calculated switching power per result bit from the original
dynamic power measurements gives \Cref{fig:heatmap2}. This reveals a second
pattern of vertical and horizontal striping that was previously obscured by the
dominant effects of the Hamming weight of the output value of the \texttt{add}
operation. Intuitively, this corresponds to the Hamming weight of both input
operands, demonstrated in \Cref{fig:inputhammingweight}.  We determine this to
be \SI{1.3}{\milli\watt} per input bit set.
The input Hamming weight predicts power in \Cref{fig:heatmap2}
with a mean error of less than 1mW.
Repeating this process and
subtracting the calculated power per input bit gives \Cref{fig:heatmap3}, which
closely corresponds to the pattern observed for the Hamming weight shown in
\Cref{fig:sumhammingweight}, as previously stated.

Finally, by subtracting the dynamic power of both the input and output bits
produces \Cref{fig:heatmap4}, which shows that the remaining variation in
dynamic power is an order of magnitude lower than the effect of these Hamming
weights, ranging from \SI{12}{\milli\watt} to \SI{0}{\milli\watt}. Expressed as a
series of matrix operations, where $P$ is the measured dynamic power and the
input and output Hamming weights are presented as $H_i$ and $H_o$ respectively,
the remaining unaccounted for dynamic power, $D$ of \Cref{fig:heatmap4}, is:
\begin{equation}
  D = P - \left(H_i \cdot 1.3\right) - \left(H_o \cdot 4.4\right) \quad
  \SI{}{\milli\watt}.
  \label{eq:powermatrix}
\end{equation}

In a real-world program, instructions would not be interleaved with the loading
of zero-value operands, as was done in our experiments. However, with each new
instruction, a Hamming distance would be present between the previous and
current input values, as well as the previous and current output value. As
such, the properties described here naturally translate from Hamming weights
into Hamming distances.

\subsubsection{Dynamic power range due to operand values}

For the \texttt{add} instruction, $P_{dmin} = \SI{34}{\milli\watt}$ and
$P_{dmax} = \SI{96}{\milli\watt}$, giving $P_{drng} = \SI{62}{\milli\watt}$.
Using \cref{eq:pctrange}, this demonstrates that for \texttt{add}, a range of
\SI{17}{}--\SI{37}{\percent} of the core's power dissipation is governed by
operand values. This is a twenty percentage point range due to operand values.
In a system where processor power is significant, this is a substantial
variation, inaccurate predictions of which may be undesirable or unsafe.

Across all of our experiments, the $P_{dmax}$ observed was
\SI{123}{\milli\watt}, caused by the \texttt{sub} instruction. This is due to
\texttt{sub} producing a negative two's complement output that results in all
bits being set in the output operand, causing maximal Hamming distance in the
output datapath. However, compared to $P_{tsingle}$ this means that on the
Xcore, dynamic switching contributes as much as \SI{42}{\percent} of the total
processor power, and that over two-thirds of this contribution is due to
operand values. Similar work for 8-bit AVR~\cite{twcem} shows dynamic power
making up \SI{15}{\percent} of processor power, which is lower than the Xcore,
most likely due to its narrower data width but also due to a range of other
architectural and manufacturing differences. The AVR, like the Xcore, is
cache-less and a candidate for use in deeply embedded systems.  Further
processor examples include variants of ARM cores in the Cortex-M series, which
do not typically include complex memory subsystems~\cite{ARMMcacheless}.

This data demonstrates that, at least on the Xcore and the AVR, the
contribution of dynamic power to the full processor cost is non-trivial, and
certainly a significant contributor to calculating the worst-case energy in
deeply embedded programs. We also observe that the result values for our
particular processor are the most significant contributor to dynamic power.
This is still input data dependent, because the result value is a
transformation of an instruction's input. For simplicity we focus only on the
most significant component, the result value, in subsequent sections:
specifically, the Hamming distance between the output values of operations
across subsequent clock cycles. We briefly discuss other components
in~\Cref{sec:discussion}.

\subsection{Summary and Discussion}

With regard to prior work that analyses the significance of dynamic power in
software execution, we have demonstrated that on the Xcore dynamic power can be
a large proportion of overall energy consumption by the processor, but cannot
discount prior work that found little contribution on other platforms.  This
suggests that dynamic power contribution \textit{can} be significant, but that
it varies from processor to processor. It should also be considered that in
real programs, sequences of instructions are unlikely to yield operand values
that produce the worst-case for each instruction. In other work \cite{twcem}
we have found energy consumption can vary up to 7\% in two benchmarks on Xcore
depending on their inputs, and up to 9\% for the same benchmarks on AVR.
However, our intent is to
examine the feasibility of finding a bound that is both safe and tight,
avoiding methods with less than absolute certainty.

The system context should also be considered, for two main reasons. Firstly, a
large system, for example that features a display and backlight component, will
have its total energy consumption dominated by these over all other
components~\cite{Carroll_Heiser_10}. Looking beyond embedded systems, large
multi-core processors such as the Xeon Phi~\cite{phimodel} consume
significantly more energy in caches and memories than in computation. This will
of course significantly reduce the impact of any variation in processor core's
energy.  Secondly, the type of system and its performance requirements will
influence processor choice, and the amount of power variation of the chosen
processor will determine whether it is necessary to consider it. If this is the
case, the computational workload placed upon the system will then determine how
much each part of the processor is exercised. It is shown in
\cite{Hameed2010a} that both processor choice and workload change how
processor subcomponents such as the register file and functional units
contribute to total energy consumption.

With this in mind, we observe that consideration of dynamic power caused by
operand values is most relevant for real-time, deeply embedded applications.
Such applications typically have energy budgets as a primary concern, have some
non-trivial processing task that requires a microcontroller, but do not use a
large processor featuring caches and other performance enhancing hardware that
would adversely affect timing predictability. When attempting to meet design
constraints such as battery lifetime, determining the worst-case energy
consumption of software would be of interest, and thus determining the impact
of data operands on dynamic energy consumption.

\section{Formalising the circuit switching problem}
\label{sec:formproblem}

As illustrated in the previous section, the matter we consider is the amount
of energy caused by circuit switching, specifically the switching occurring on
the output datapath in a processor. Here, we
formalise our problem, which we name the ``Circuit SWitching
Problem'' (CSWP), discussing its limitations and generality. Our objective
is to take a program, determine the maximum amount
of output datapath switching activity that can occur in that program, and, in
the process, find the program input that triggers it.

Because we are only concerned with the amount of circuit switching that can
occur due to operand values, we choose to limit the problem and avoid any
facility for varying the length of a program in this formalisation, i.e. the
number of instructions executed.  A CSWP program thus cannot have any branch
instructions or conditional execution ability: it corresponds closely with a
trace of a general program execution, or a general program that has been
unrolled and all conditional branches eliminated. Dealing with programs of
varying length would involve searching different paths through the program, and
the variation of energy consumed by static power would become a significant
consideration.  We chose to work with fixed-length CSWP programs to maintain
focus on switching in data operands.

Formally, we consider a CSWP program, $X$, to be a finite sequence of $n$
instructions, $x_i$, such that $X = {x_1, x_2, ..., x_n}$. Each instruction is
a 3-tuple $\langle m, i, o \rangle$, where $m$ is a mnemonic $m \in M$, $i$ is
a set of input operands (discussed below), and $o$ is an output operand. Both
inputs and outputs (discussed further below) are considered to be bit-vectors
of width $w$.

A CSWP program executes on an abstract machine with a monotonically
incrementing program counter, an infinite number of registers, and a memory
store of finite size. Memory is considered to be an array of size $2^w$ with
each memory cell a bit-vector of width $w$. For each instruction $x_i$ in the
CSWP program the machine takes the input operands, computes an output according
to the function of the instruction mnemonic, and writes the result to the
output operand. The objective function of CSWP is then to compute:
\begin{equation}
	\sum^{n-1}_{i=1} h(o_i, o_{i+1})\ ,
\end{equation}
where $h$ is a function computing Hamming distance between two values, i.e.
the output values of each subsequent instruction, corresponding to the output datapath
of the abstract machine.

Each mnemonic $m \in M$ represents a function over the input operands,
resulting in a single output result. In line with the constraints detailed
above, CSWP programs only perform arithmetic and logic computations, mapping input
operands to an output. There are no branch mnemonics, neither are there any
instructions that induce side effects of any form (such as changing some state
or the program counter).  We do not define a set of mnemonics that a CSWP
program may use, however for the purposes of this paper we write listings using
standard RISC mnemonics such as \texttt{add}, \texttt{sub}, \texttt{ldr},
\texttt{mov}~\cite{comparch}.

Each input operand is permitted to be one of four classes of sources:
\begin{itemize}
	\item Free inputs, which we denote with the text \textit{free}.
	\item Constant values, which we write in hexadecimal, e.g. \texttt{\em 0x1}.
	\item A memory access to a fixed address $m[x]$, with $x$ the address.
	\item The output result of a prior instruction, written $o_i$, where for the current instruction $x_j$, $i < j$.
\end{itemize}

The value of every input is always a bit-vector of width $w$. Free inputs may
take any value, likewise constants may only have one value, defined in the
instruction being executed. Memory accesses evaluate to the contents of a
memory cell, but for simplicity we only permit the addressing of fixed memory
addresses. Prior output operands correspond to the output of each instruction
being written to one of the registers,
which may then be read as an input to another instruction.

All instructions are considered to have an output of bit-width $w$, i.e., they
all write some value to the output datapath of the machine. A \texttt{nop}
(no-operation) instruction would be any instruction that repeats the output value
of the previous instruction, causing no switching activity on the output datapath.
Outputs may
optionally be written to a memory cell $m[x]$, where $x$ is a fixed address for
the output value to be written to. In this circumstance, the output value may
still be referred to as $o_i$, as a store to memory
still causes the bits in the machine's result datapath to flip.

This formalisation has a number of limitations, most notably that without an
infinite data store or ability to programmatically address it, it is not Turing
complete. Given that our aim is to find the maximum switching for a particular
path through a general program, this is a
suitable restriction. The formalisation does not correspond to a particular
machine, although with additional restrictions it may correctly model the
execution trace of existing processors. The memory array may be considered to
be superfluous given the lack of complex addressing, however it provides a
useful mechanism for illustrating our examples through the rest of this paper.

We observe that CSWP is in class NP, as one may easily check the validity of a
solution. Given the CSWP program and an input valuation for each free input,
we can simulate the program with the given inputs, counting the number of bit
flips at the same time. The complexity of this process scales linearly
with the number of instructions, $n$.

\section{Reducing MAXSAT2 to the circuit switching problem}
\label{sec:frommaxsat}

To demonstrate that the CSWP is NP-hard, we must reduce any NP-hard problem
to CSWP in polynomial time. For this, we turn to the MAXSAT problem,
which is known to be NP-hard~\cite{sathandbook}. Specifically, we work with the
MAXSAT2 variant, where each clause is limited to having at most two literals.
Despite 2SAT being solvable in polynomial time, MAXSAT2 is still known to be
NP-hard~\cite{approxalgs}.

We reduce MAXSAT2 to CSWP by simulating MAXSAT2 in the switching activity of
an instruction sequence, where the input that causes the maximum amount of
circuit switching corresonds to an assignment to the Boolean variables that
causes the maximum number of clauses to be satisfied. The reduction is
illustrated in \Cref{alg:reduce}, which takes the number of Boolean variables
and the set of clauses as input, and outputs a CSWP program that
simulates MAXSAT2. Here, we assume that the function \texttt{PrintInsn} causes
a CSWP instruction to be emitted from the algorithm, with the instruction
mnemonic, set of variables, and optional output destination as its respective
arguments. The return value identifies the output operand of the instruction.

\begin{algorithm}
	\SetKwFunction{PrintInsn}{PrintInsn}
	\SetKwFunction{LitToMemAddr}{LitToMemAddr}
	\KwIn{Number of variables $n$ and set of clauses $C$}
	\KwOut{CSWP program encoding MAXSAT2 problem}
	var\_addr = \texttt{0}\;
	\For{i = \texttt{0} $;$ i $< n$ $;$ i++} {
		out1 = \PrintInsn{``mov'', [free]}\;
		out2 = \PrintInsn{``xor'', [out1, \texttt{0x1}]}\;
		\PrintInsn{``store'', [out1], m[var\_addr++]}\;
		\PrintInsn{``store'', [out2], m[var\_addr++]}\;
		\PrintInsn{``mov'', [\texttt{0}]}\;
	}
	\ForEach{$c \in C$}{
		$<l1, l2> = c$\;
		laddr1 = \LitToMemAddr{l1}\;
		laddr2 = \LitToMemAddr{l2}\;
		lit1 = \PrintInsn{``load'', [m[laddr1]]}\;
		\PrintInsn{``xor'', [lit1, \texttt{0x1}]}\;
		\PrintInsn{``mov'', [\texttt{0}]}\;
		lit2 = \PrintInsn{``load'', [m[laddr2]]}\;
		\PrintInsn{``xor'', [lit2, \texttt{0x1}}\;
		\PrintInsn{``mov'', [\texttt{0}]}\;
		\PrintInsn{``or'', [lit1, lit2]}\;
		\PrintInsn{``mov'', [\texttt{0}]}\;
	}
\caption{Algorithm for encoding of MAXSAT2 formula within a CSWP program, printed via PrintInsn.}
\label{alg:reduce}
\end{algorithm}

First, we read a series of free input values, and for the moment we assume they lie in the range
$[0, 1]$, i.e. represent true or false in the lowest bit of the otherwise zero bit-vector. We
consider each of these values to be an assignment to a Boolean variable in the
MAXSAT2 problem. Each bit, and its complement, are stored to a location in
memory. This creates an array of values corresponding to the truth of each
literal. At the end of this process we
insert a \texttt{mov} instruction that loads a zero value, for the purpose of
resetting the value on the output datapath to zero.
The net effect is that for each Boolean variable read, a constant amount of
switching activity occurs. Consider each value the free variable may have:
\begin{enumerate}
	\item \texttt{True}: Reading the input switches the lowest output
		datapath bit to on, the subsequent \texttt{xor} switches it to
		off, and the final \texttt{mov} causes no switching.
	\item \texttt{False}: Reading the input causes no switching, the
		\texttt{xor} switches the lowest output datapath bit to on,
		and the subsequent \texttt{mov} switches it back to off.
\end{enumerate}
Thus, for each Boolean variable read, the CSWP program always causes two bit
flips.

We then proceed to use the memory region prepared with literal valuations to
simulate the MAXSAT2 problem. We assume a mapping between each literal of the
Boolean variables and the address of its valuation in the memory array, and
use the function \texttt{LitToMemAddr} to translate from literal to memory
address. Then, for each clause, we produce an instruction sequence
that loads each literal valuation using the constant-switching technique
used to read free inputs. Once the literals are loaded, they are \texttt{or}'d
together, after which the output datapath is loaded with zero again.

The CSWP program produced by \Cref{alg:reduce} has both a constant and
data dependent  portion of switching activity. Two bit-flips occur for each Boolean
variable in the input MAXSAT2 problem, and four for each clause. The switching
activity from the \texttt{or} instruction, however, directly corresponds to the
satisfiability of the clauses: if a clause is satisfiable (i.e., one
of the literals is true) then the \texttt{or} and following \texttt{mov} will
cause two additional bit-flips. If a clause is not satisfiable, the same
instructions will cause no switching. As a result, the maximum amount of
switching in the program is caused by the maximum number of clauses
being satisfied. The assignment to the free variables which causes this is
also an assignment to the Boolean variables of the MAXSAT2 problem that
causes the maximum number of clauses to be satisfied. As a result,
CSWP must be at least as hard as MAXSAT2 (i.e. NP-hard). As we know CSWP is also
in class NP (\Cref{sec:formproblem}), CWSP is NP-hard. $\square$

The assumption that free variables are either one or zero is to simplify the
presentation: a similar technique to balance switching can be used to translate
any input word into one or zero, with constant switching. For example on a
machine with a condition status register, one may compare the input register
with zero, exclusive-or the entire input register with an all-ones value,
then store the comparison result flag to a general register and continue with the
procedure above.

We observe that the reduction is performed in polynomial time, as it scales
linearly with the number of Boolean variables $n$ and the number of
clauses, of which there can be at most $n^2$.

Given this result, we can conclude that there cannot be an efficient algorithm
that solves the CSWP, unless P $=$ NP. Thus, given that general programs can be
unrolled and reduced to a CSWP, it is infeasible to determine the
\textit{exact} worst-case switching due to operand values in a program, defeating energy estimation techniques that
would rely on such a model. This corresponds to the first form of WCEC
estimate given in \Cref{sec:wcecbackground}.
However, given such a limitation, there could
still be algorithms that approximate the worst-case switching to a certain
degree of accuracy, allowing worst-case switching to be narrowed down to a
small range of values. We address this in the next section.

\section{Inapproximability}
\label{sec:inapprox}

\begin{algorithm}
	\SetKwFunction{CheckSat}{CheckSat}
	\KwIn{Number of variables $n$ and set of clauses $C$}
	\KwOut{CSWP program with switching gap}
	\tcc{Decision phase}
	base\_var\_addr = var\_addr = \texttt{0}\;
	insn\_count = \texttt{0}\;
	\For{i = 0 $;$ i $< n$ $;$ i++} {
		out1 = \PrintInsn{``mov'', [free]}\;
		\PrintInsn{``store'', [out1], m[var\_addr++]}\;
	}
	result = \CheckSat{base\_var\_addr, $C$}\;
	bit\_pattern = \PrintInsn{``ite'', [result, \texttt{0xFFFFFFFF}, \texttt{0}]}\;
	\tcc{Switching phase}
	decision\_insn\_count = insn\_count\;
	\For{i = \texttt{0} $;$ i $< ((decision\_insn\_count)/\texttt{2} + \texttt{1})$ $;$ i++} {
		\PrintInsn{``mov'', [bit\_pattern]}\;
		\PrintInsn{``mov'', [\texttt{0}]}\;
	}
\caption{Algorithm encoding a SAT problem into CSWP, with an output gap governed by satisfiability}
\label{alg:approx}
\end{algorithm}

Having shown that CSWP is NP-hard, we will now show that it also cannot be
approximated to any useful factor, i.e. the second form of WCEC estimate in
\Cref{sec:wcecbackground} is infeasible. We demonstrate that there is no constant
$\varepsilon$ for which an approximation factor of $1 - \varepsilon$ can be
achieved, and then that polynomial approximation factors also cannot be
achieved. Intuitively, this is because each bit flip caused by the program is
the result of an arbitrary computation, meaning there is no structure to the
combinatorial problem that one can generally rely upon when constructing an
approximation.

Formally, we demonstrate CSWPs inapproximability using a gap introducing
reduction~\cite{approxalgs} from SAT to CSWP. Such a reduction transforms an
NP-complete decision problem into an NP-hard optimisation problem, with a quantity (the
``gap'') of the feature being optimised governed by the truth of the decision
problem. By demonstrating such a gap, one shows that a portion of the
NP-hard problem cannot be approximated in polynomial time, as the approximation
algorithm would have to solve an NP-complete problem in the process.

In the context of CSWP, we demonstrate that for any instance of SAT problem
$p$, we can reduce it to a CSWP program $q$ where a portion of the switching
activity is governed by the truth of whether $p$ is satisfiable. The
transformation is illustrated in \Cref{alg:approx}, which we divide into two
discrete portions: the decision phase, and the switching phase. We use the same
functions as in \Cref{alg:reduce}, with the modification that the
\texttt{PrintInsn} function increments a counter, \texttt{insn\_count},
for every instruction printed.

Throughout the decision phase, we are not concerned with the switching activity
that may occur, and do not seek to control it, in contrast with the previous algorithm.
We begin by reading $n$ free variables, which for now we assume to be bit-vectors with
either zero or one in the least significant bit and all other bits zero.
These free variables are stored to fixed addresses in memory. We then pass
the address of the variable valuations and the SAT clauses to the
\texttt{CheckSat} function, which emits a CSWP program that evaluates the
clauses over the Boolean variables stored at \texttt{base\_var\_addr}, and returns an
output operand identifying whether the assignment satisfied the clauses.
Significantly, we do not seek to define how \texttt{CheckSat} checks the
satisfiability of the clauses, we only assume that it achieves it in a number
of instructions polynomial in $n$, the number of Boolean variables. We know
that SAT is in NP, so due to complexity theory we also know an assignment can
be verified in a polynomial number of instructions.\footnote{We note that, as
the inputs to
\texttt{CheckSat} are free variables, we are essentially modelling a SAT
solver.} We then produce an output, \texttt{bit\_pattern}, using an
``if-then-else'' instruction that evaluates to zero if the Boolean variables
do not satisfy the clauses, and has all bits set if they do.

For the switching phase,
the CSWP instruction counter, \texttt{insn\_count}, is read to learn how
many instructions there are in the decision phase of the CSWP program. We then
emit a pattern that repeatedly loads the variable \texttt{bit\_pattern} and
then zero. The effect of this is to produce a phase in the program
that causes a large amount of switching if the SAT problem $p$ was satisfied;
and to not if it was unsatisfiable. In this sequence, a satisfying assignment will cause
the switching phase to flip every bit in the result datapath, every
instruction; while no switching will occur otherwise.

We have thus introduced a gap in the switching activity of the CSWP program
$q$, that is governed by whether the SAT problem $p$ is satisfiable or not.
We use the length of the decision phase of the program to ensure that the
switching phase is at least the length of the decision phase, plus one or two
instructions. This ensures that, regardless of the amount of switching in the
decision phase, the switching phase dominates the switching activity of the
program. When solving CSWP, if the SAT problem $p$ were satisfiable, then the
maximum amount of switching would include the switching phase, and the CSWP
solver would be obliged to yield an input to the program that satisfied the
reduced SAT problem. If $p$ is unsatisfiable, it would instead yield whatever
input maximised the switching in the decision phase.

We use the size of the gap to demonstrate that CSWP cannot be approximated.
In the previous example the switching phase constitutes at least $1/2$ of the possible switching
activity: if one possessed an algorithm to approximate such a CSWP program to within a
factor of $1/2$, then it would be obliged to activate the switching phase of
any CSWP program reduced from a satisfiable SAT formula, thus acting as an
oracle for an NP-complete problem. Under the P $\ne$ NP assumption, such an
algorithm does not exist. $\square$

Furthermore, we are able to extend this result to any constant factor.  For any
value of $\varepsilon$ and SAT instance $p$, take the desired approximation
factor $f = 1 - \varepsilon$ and set the length of the switching phase to be
$declen \times (1/f)$, where $declen$ is the number of instructions in the decision
phase of CSWP $q$. Such a program will have a gap of at least $1/f$ times the
decision phase, that depends entirely on the satisfiability of $p$, and thus
cannot be approximated. One need not limit this approach to a constant factor
either: one may instead compute $f$ to be some factor that is a polynomial
function of the size of SAT problem $p$, for example $n^2$, and achieve the
same result. This shows that there can be no useful approximation factor for
CSWP.

The safety of this result depends on the reduction to $q$ being polynomial in
the number of variables $n$ in $p$.
Introducing the variables of $p$ scales linearly with $n$, checking
the satisfiability of a particular assignment is known to be checkable in
polynomial time, and the evaluation of the result into \texttt{bit\_pattern}
is constant-time. The decision phase is thus a polynomial reduction.
The switching phase is controlled by the length of the decision phase
(which is polynomial), but also the desired approximation factor. Provided
the approximation factor is polynomial, the full reduction is also polynomial.

The assumption in \Cref{alg:approx} that free inputs are only zero or one is again to ease
presentation: the precise value is irrelevant so long as the switching phase
accounts for the maximum amount of switching it can cause. The only requirement
is that a sufficient number of free bits are supplied to \texttt{CheckSat} to represent
the free inputs to the SAT problem $p$.

\section{Discussion}
\label{sec:discussion}

We consider here the scope of these results with regards to the hardware for which
dynamic operand energy should be a consideration, and the
implications of these results for WCEC analysis.

\subsection{Hardware scope}

Determining a program's switching activity caused by operand
values is NP-hard,
therefore calculating the worst-case dynamic energy
for a program in a way that accounts for its input data set is infeasible.
Our result is relevant when the cost of dynamic power is dominated
by switching in the output datapath.
Clearly, the exact cost of such switching will vary
between processors, however our result may be used as a basis for demonstrating
that calculating the switching in other components of the processor is also
infeasible. For example, because all inputs to instructions are inevitably
a direct input or the
result of some other instruction, it is reasonable to assume
that it is NP-hard to estimate the switching activity of input operands too.

\subsubsection{Processor architecture considerations}

The simplest processor architectures, such as deeply embedded AVR or ARM M0
that might be found in low power IoT scenarios, feature few additional
components that would make further contributions to dynamic energy consumption.
However, more complex architectures, present in larger IoT and embedded
systems, introduce extra contributions.

For example, data caches will contribute dynamic energy too. First, in the
computations they perform to decide upon an outcome, for example determining if
a memory address is present in cache. Second is the cost of that outcome, for
example the subsequent request to a higher-level cache or main memory. These
also depend on program inputs to an extent, but are not modelled by our CSWP
formalisation.  Other processor components may contribute dynamic energy that
is not affected by the inputs to a program. The switching associated with
instruction logic (decode, functional unit activation, instruction cache) may
contribute dynamic energy regardless of the program input.

Features in some processors, such as out-of-order execution may also defeat our
analysis. The circuit switching cost is still present, and its determination
will still be NP-hard, however it may occur in an unpredictable fashion that
depends on a processor-internal unobservable instruction execution schedule.

Energy saving mechanisms such as power- and clock-gating may also reduce the
impact of dynamic energy due to operand values. However, if present, the extent
of their effect will be dependent upon the architecture design as well as the
technology with which it is manufactured. Coarse-grained mechanisms such as
sleep states may also warrant attention, although these may need to be
accounted for differently if they are directly controlled by software.

\subsubsection{System-level considerations}

Considering again a cache hierarchy, difference in its dynamic power
contribution due to the cost of a hit versus the cost of a miss is likely to be
higher than the range of dynamic power due to operand values within the
processor core. If this cache cost is added to the dynamic energy of the
processor core, then the dynamic variation due to operand values is now a
much smaller proportion of total energy consumption.

In such a system, it remains infeasible to determine a worst-case due to
operand values as per our CSWP formalisation. While the impact of this may be
lower, there remains a level of uncertainty that risks compromising the safety
of any energy consumption assumptions that are made when modelling is
performed. A mitigation strategy, such as more conservative assumptions of
worst-case energy per instruction, must therefore be used. The accuracy of
cache modelling must also now be considered, where safety of any energy bounds
is compromised if the model does not precisely reflect cache hits and misses
for a program.

Beyond caches, systems may have peripherals that eclipse the processor in terms
of energy consumption, such as wireless transmitters for communication-heavy
IoT applications. In such
circumstances, the impact of unsafe or overly-conservative worst-case
instruction energy modelling is unlikely to be of concern in the context of
energy budgeting.

In summary, it is essential to assess the proportion of energy consumed in the system by
the processor, and its dynamic consumption due to operand value, to determine
if CSWP should be a concern. This is primarily why the focus of this paper has
been upon processors for deeply embedded systems.

\subsection{Implications for WCEC analysis}

\subsubsection{No guarantees}

The infeasibility result for estimating dynamic operand energy over time
renders the first two forms of WCEC estimation discussed in \Cref{sec:wcecbackground}
infeasible. Further, it prevents the
construction of an instruction level energy model that identifies an accurate
worst-case switching cost for each instruction in a given program.
Existing techniques that apply WCEC analysis~\cite{wcec-jayaseelan,lopstr13,scopes15,fopara15,tacopapertr,wcec-wagemann}
to software can thus never have an energy model that accurately accounts for
worst-case achievable dynamic energy of the given computation.

We are still left with the coarse over-approximation WCEC estimation techniques.
Such techniques provide a safe upper bound but no relationship between that
bound and the true WCEC, and our result shows that any such relationship would
be infeasible to calculate.
For example, calculating the absolute maximum operand switching cost for an
execution in the manner of~\cite{wcec-jayaseelan} and~\cite{wcec-wagemann}
would be sufficient.
The over-approximation inherent with this approach will not yield a tight
bound. For example, on the XMOS XS1-L, with dynamic energy contributing
\SI{42}{\percent} of energy consumption, one would have a similarly sized
amount of potential over-approximation regarding the energy consumption of any
execution. The expected over-approximation would therefore be somewhere within this range, likely towards the middle if an average case is assumed.

\subsubsection{Alternatives}

Viable techniques for estimating dynamic energy consumption can come from a
variety of fields: in particular, statistical methods~\cite{twcem} may be
effective for determining the distribution of energy consumption under normal
operation. Such a model may be used by assuming that the most energy the
program can consume occurs only \SI{1}{\percent} of the time, and taking the
energy value corresponding to that probability as the program's worst-case energy
consumption. This does not provide a safe upper bound on the program's energy
consumption as it is based on normal operation. However, on the balance of
probability it is very likely to present an upper bound. Depending on the use
case, such a bound may be more useful in making energy consumption of software
transparent to developers than gross over-approximation.

Another alternative is to refine coarse over-approximations: simply assuming
maximum switching activity for the whole length of the program yields a likely
very inaccurate upper bound. Further techniques such as static analysis or
abstract interpretation could reduce this inaccuracy. For example, if one can
determine the integer interval of a variable, then one can potentially
bound the amount of switching between adjacent instructions, and thus
determine the maximum switching of a specific instruction sequence to be lower
than its absolute maximum.

In all circumstances, alternative estimation techniques will posses some level
of unquantifiable over-approximation, unsafeness or incompleteness,
otherwise they will be NP-hard as proven in this work.

\section{Conclusions and future work}
\label{sec:conclusions}

In this paper we have considered the energy consumption in a processor that can
directly be attributed to the data or inputs to the software being executed,
and demonstrate that the general analysis of circuit switching in processor
datapaths --- the ``circuit switching problem'' --- is NP-hard. Further, we
demonstrate that there is no efficient approximation algorithm for the circuit
switching problem to any constant or polynomial factor. We conclude that
only the coarsest of estimation techniques can be used in the analysis of
worst-case energy in polynomial time. This limitation introduces an uncertain
amount of over-approximation in the gap between the true WCEC and the
estimated WCEC.

We consider alternate techniques and questions that one could pose that do not
amount to worst-case analysis but do provide an estimate of how large the worst
case could be, and how they can contribute to understanding software
energy consumption.

In the future we believe that work is best focused on statistical methods of
modelling program energy consumption, or otherwise characterising the way in
which software operates.
Similar efforts are being made that model the WCET problem probabilistically \cite{pwcet},
for example using extreme value theory (EVT) \cite{longin2016extreme},
although recent research suggests EVT may not be completely applicable
to WCET in general \cite{pwcet-a-careful-look}.
Critically, we cannot continue to think in terms of a mathematically proven
``worst-case'' energy consumption, but must instead turn to other methods for
energy consumption analysis that may not be sound or accurate, but are at least
feasible.

Further exploration of the scope of data-dependent energy would also benefit
WCEC. For example, whilst it is intuitive that a cache miss has a higher energy
cost than a cache hit by a significant margin, how large a contribution to
dynamic power is made by the address and data values on the buses in the memory
hierarchy? This would establish whether caches can present energy anomalies of
a similar nature to already studied timing anomalies. In addition, this could
further address how tight a safe WCEC bound can be in more complex
microarchitectures.

\subsection{Acknowledgements}
We would like to thank David May, Benjamin Sach, Kyriakos Georgiou and James Pallister
for their insights into and motivation of this work. The research leading to
these results has received funding from the European Union 7th Framework
Programme (FP7/2007-2013) under grant agreement no 318337, ENTRA -
Whole-Systems Energy Transparency; and grant agreement no 611004,
ICT-Energy.

\bibliographystyle{plain}
\bibliography{theory}

\begin{thebibliography}{10}

\bibitem{ARMMcacheless}
ARM.
\newblock Arm cortex-m programming guide to memory barrier instructions.
\newblock Technical report, ARM, 2012.

\bibitem{highdatadep}
Giuseppe Ascia, Vincenzo Catania, Maurizio Palesi, and Davide Sarta.
\newblock {An instruction-level power analysis model with data dependency}.
\newblock {\em {VLSI DESIGN}}, {12}({2}):{245--273}, {2001}.

\bibitem{sathandbook}
Armin Biere, Marijn Heule, Hans van Maaren, and Toby Walsh.
\newblock {\em Handbook of Satisfiability: Volume 185 Frontiers in Artificial
  Intelligence and Applications}.
\newblock IOS Press, Amsterdam, The Netherlands, The Netherlands, 2009.

\bibitem{butzen2006leakage}
Paulo~Francisco Butzen and Renato~Perez Ribas.
\newblock Leakage current in sub-micrometer cmos gates.
\newblock Technical report, Universidade Federal do Rio Grande do Sul, 2006.

\bibitem{Carroll_Heiser_10}
Aaron Carroll and Gernot Heiser.
\newblock {An analysis of power consumption in a smartphone}.
\newblock In {\em Proceedings of the 2010 USENIX conference on USENIX annual
  technical conference}, USENIXATC'10, page~21, Berkeley, CA, USA, 2010. USENIX
  Association.

\bibitem{pwcet}
Francisco~J. Cazorla, Tullio Vardanega, Eduardo Qui{\~n}ones, and Jaume Abella.
\newblock {Upper-bounding Program Execution Time with Extreme Value Theory}.
\newblock In Claire Maiza, editor, {\em 13th International Workshop on
  Worst-Case Execution Time Analysis}, volume~30 of {\em OpenAccess Series in
  Informatics (OASIcs)}, pages 64--76, Dagstuhl, Germany, 2013. Schloss
  Dagstuhl--Leibniz-Zentrum fuer Informatik.

\bibitem{Chandra2008}
Lokesh Chandra and Sourav Roy.
\newblock {Estimation of energy consumed by software in processor caches}.
\newblock In {\em 2008 IEEE International Symposium on VLSI Design, Automation
  and Test (VLSI-DAT)}, pages 21--24. IEEE, April 2008.

\bibitem{onthecomplexity}
Ana~T. Freitas, Horácio~C. Neto, and Arlindo~L. Oliveira.
\newblock {On the cmoplexity of Power Estimation Problems}.
\newblock 2004.

\bibitem{tacopapertr}
Kyriakos Georgiou, Steve Kerrison, and Kerstin Eder.
\newblock On the value and limits of multi-level energy consumption static
  analysis for deeply embedded single and multi-threaded programs.
\newblock Technical report, University of Bristol, 2015.

\bibitem{gonzalez1997supply}
Ricardo Gonzalez, Benjamin~M Gordon, and Mark~A Horowitz.
\newblock Supply and threshold voltage scaling for low power cmos.
\newblock {\em IEEE Journal of Solid-State Circuits}, 32(8):1210--1216, 1997.

\bibitem{scopes15}
Neville Grech, Kyriakos Georgiou, James Pallister, Steve Kerrison, Jeremy
  Morse, and Kerstin Eder.
\newblock Static analysis of energy consumption for llvm ir programs.
\newblock In {\em Proceedings of the 18th International Workshop on Software
  and Compilers for Embedded Systems}, SCOPES '15, pages 12--21, New York, NY,
  USA, 2015. ACM.

\bibitem{peakpower}
Hadi Hajimiri, Kamran Rahmani, and Prabhat Mishra.
\newblock Efficient peak power estimation using probabilistic cost-benefit
  analysis.
\newblock In {\em VLSI Design (VLSID), 2015 28th International Conference on},
  pages 369--374, Jan 2015.

\bibitem{Hameed2010a}
Rehan Hameed, Wajahat Qadeer, Megan Wachs, Omid Azizi, Alex Solomatnikov,
  Benjamin~C. Lee, Stephen Richardson, Christos Kozyrakis, and Mark Horowitz.
\newblock {Understanding sources of inefficiency in general-purpose chips}.
\newblock {\em Proceedings of the 37th annual international symposium on
  Computer architecture - ISCA '10}, page~37, 2010.

\bibitem{comparch}
John~L. Hennessy and David~A. Patterson.
\newblock {\em Computer Architecture, Fifth Edition: A Quantitative Approach}.
\newblock Morgan Kaufmann Publishers Inc., San Francisco, CA, USA, 5th edition,
  2011.

\bibitem{wcec-jayaseelan}
Ramkumar Jayaseelan, Tulika Mitra, and Xianfeng Li.
\newblock Estimating the worst-case energy consumption of embedded software.
\newblock In {\em Proceedings of the 12th IEEE Real-Time and Embedded
  Technology and Applications Symposium}, RTAS '06, pages 81--90, Washington,
  DC, USA, 2006. IEEE Computer Society.

\bibitem{Johnson}
David~S. Johnson.
\newblock Approximation algorithms for combinatorial problems.
\newblock In {\em Proceedings of the Fifth Annual ACM Symposium on Theory of
  Computing}, STOC '73, pages 38--49, New York, NY, USA, 1973. ACM.

\bibitem{async-timing}
David Kearney and Neil~W. Bergmann.
\newblock Performance evaluation of asynchronous logic pipelines with data
  dependent processing delays.
\newblock In {\em Asynchronous Design Methodologies, 1995. Proceedings., Second
  Working Conference on}, pages 4--13, May 1995.

\bibitem{steve-tecs}
Steve Kerrison and Kerstin Eder.
\newblock Energy modeling of software for a hardware multithreaded embedded
  microprocessor.
\newblock {\em {ACM} Trans. Embedded Comput. Syst.}, 14(3):56, 2015.

\bibitem{kim2003leakage}
NS~Kim, T~Austin, D~Baauw, and T~Mudge.
\newblock {Leakage current: Moore's law meets static power}.
\newblock {\em Computer}, pages 68--75, 2003.

\bibitem{Lee2011}
Yebin Lee and Soontae Kim.
\newblock {DRAM energy reduction by prefetching-based memory traffic
  clustering}.
\newblock {\em Proceedings of the 21st edition of the great lakes symposium on
  Great lakes symposium on VLSI - GLSVLSI '11}, page 103, 2011.

\bibitem{pwcet-a-careful-look}
G.~Lima, D.~Dias, and E.~Barros.
\newblock Extreme value theory for estimating task execution time bounds: A
  careful look.
\newblock In {\em 2016 28th Euromicro Conference on Real-Time Systems (ECRTS)},
  pages 200--211, July 2016.

\bibitem{fopara15}
Umer. Liqat, Kyriakos. Georgiou, Steve. Kerrison, Pedro. Lopez-Garcia, John~P.
  Gallagher, Manuel~V. Hermenegildo, and Kerstin. Eder.
\newblock {\em Inferring Parametric Energy Consumption Functions at Different
  Software Levels: ISA vs. LLVM IR}, pages 81--100.
\newblock Springer International Publishing, Cham, 2016.

\bibitem{lopstr13}
Umer Liqat, Steve Kerrison, Alejandro Serrano, Kyriakos Georgiou, Pedro
  Lopez-Garcia, Neville Grech, Manuel~V. Hermenegildo, and Kerstin Eder.
\newblock {E}nergy {C}onsumption {A}nalysis of {P}rograms based on {XMOS}
  {ISA}-level {M}odels.
\newblock In {\em Logic-Based Program Synthesis and Transformation, 23rd
  International Symposium, {LOPSTR} 2013, Revised Selected Papers}, volume 8901
  of {\em Lecture Notes in Computer Science}, pages 72--90. Springer, 2014.

\bibitem{longin2016extreme}
F.~Longin.
\newblock {\em Extreme Events in Finance: A Handbook of Extreme Value Theory
  and its Applications}.
\newblock Wiley Handbooks in Financial Engineering and Econometrics. Wiley,
  2016.

\bibitem{Lundqvist:1999:TAD:827271.829103}
Thomas Lundqvist and Per Stenstr\"{o}m.
\newblock Timing anomalies in dynamically scheduled microprocessors.
\newblock In {\em Proceedings of the 20th IEEE Real-Time Systems Symposium},
  RTSS '99, pages 12--, Washington, DC, USA, 1999. IEEE Computer Society.

\bibitem{XS1-Architecture}
David May.
\newblock The {XMOS} {XS1} architecture. available online:
  http://www.xmos.com/published/xmos-xs1-architecture, 2013.

\bibitem{Hsiao1997}
{Michael Hsiao et al.}
\newblock {K2: an estimator for peak sustainable power of VLSI circuits}.
\newblock {\em Low Power Electronics and Design}, 1997.

\bibitem{realisticpower}
Pedro~Marques Morgado, Paulo~F. Flores, and Luis~Miguel Silveira.
\newblock Generating realistic stimuli for accurate power grid analysis.
\newblock {\em ACM Trans. Des. Autom. Electron. Syst.}, 14(3):40:1--40:26, June
  2009.

\bibitem{Pallister2014}
James Pallister, Kerstin Eder, Simon~J. Hollis, and Jeremy Bennett.
\newblock {A high-level model of embedded flash energy consumption}.
\newblock In {\em Proceedings of the 2014 International Conference on
  Compilers, Architecture and Synthesis for Embedded Systems - CASES '14},
  pages 1--9, New York, New York, USA, 2014. ACM Press.

\bibitem{twcem}
James Pallister, Steve Kerrison, Jeremy Morse, and Kerstin Eder.
\newblock Data dependent energy modelling: {A} worst case perspective.
\newblock In {\em Proceedings of the 18th International Workshop on Software
  and Compilers for Embedded Systems}, SCOPES '17, New York, NY, USA, 2017.
  ACM.
\newblock To appear, preprint at \url{http://arxiv.org/abs/1505.03374}.

\bibitem{Parikh2004}
Amisha Parikh, Soontae Kim, Mahmut~T. Kandemir, Narayanan Vijaykrishnan, and
  Mary~Jane Irwin.
\newblock Instruction scheduling for low power.
\newblock {\em Journal of VLSI signal processing systems for signal, image and
  video technology}, 37(1):129--149, 2004.

\bibitem{highlevelmodels}
Suzanne Rivoire, Parthasarathy Ranganathan, and Christos Kozyrakis.
\newblock A comparison of high-level full-system power models.
\newblock In {\em Proceedings of the 2008 Conference on Power Aware Computing
  and Systems}, HotPower'08, pages 3--3, Berkeley, CA, USA, 2008. USENIX
  Association.

\bibitem{satbook2013}
Uwe Sch{\"o}ning and Jacobo Tor{\'a}n.
\newblock {\em The Satisfiability Problem: Algorithms and Analyses}.
\newblock Mathematik f{\"u}r Anwendungen. Lehmanns Media, 2013.

\bibitem{phimodel}
Yakun~Sophia Shao and David Brooks.
\newblock {Energy characterization and instruction-level energy model of
  Intel's Xeon Phi processor}.
\newblock In {\em International Symposium on Low Power Electronics and Design
  (ISLPED)}, number November, pages 389--394. IEEE, September 2013.

\bibitem{energyawaresoftware}
Amit Sinha and Anantha~P. Chandrakasan.
\newblock Energy aware software.
\newblock In {\em Proceedings of the 13th International Conference on VLSI
  Design}, VLSID '00, pages 50--, Washington, DC, USA, 2000. IEEE Computer
  Society.

\bibitem{Steinke2001}
Stefan Steinke, Markus Knauer, Lars Wehmeyer, and Peter Marwedel.
\newblock {An Accurate and Fine Grain Instruction-level Energy Model Supporting
  Software Optimizations}.
\newblock In {\em Proceedings of PATMOS}, 2001.

\bibitem{thiele:2004}
Lothar Thiele and Reinhard Wilhelm.
\newblock Design for timing predictability.
\newblock {\em Real-Time Syst.}, 28(2-3):157--177, November 2004.

\bibitem{lowcircuitcost}
Vivek Tiwari, Sharad Malik, and Andrew Wolfe.
\newblock Power analysis of embedded software: a first step towards software
  power minimization.
\newblock {\em Very Large Scale Integration (VLSI) Systems, IEEE Transactions
  on}, 2(4):437--445, Dec 1994.

\bibitem{tiwarimodel}
Vivek Tiwari, Sharad Malik, Andrew Wolfe, and Mike Tien-Chien Lee.
\newblock Instruction level power analysis and optimization of software.
\newblock {\em J. VLSI Signal Process. Syst.}, 13(2-3):223--238, August 1996.

\bibitem{approxalgs}
Vijay~V. Vazirani.
\newblock {\em Approximation Algorithms}.
\newblock Springer-Verlag New York, Inc., New York, NY, USA, 2001.

\bibitem{wcec-wagemann}
Peter W\"agemann, Tobias Distler, Timo H\"onig, Heiko Janker, R\"udiger
  Kapitza, and Wolfgang Schr\"oder-Preikschat.
\newblock Worst-case energy consumption analysis for energy-constrained
  embedded systems.
\newblock In {\em Real-Time Systems (ECRTS), 2015 27th Euromicro Conference
  on}, pages 105--114, July 2015.

\bibitem{wcetsurvey}
Reinhard Wilhelm, Jakob Engblom, Andreas Ermedahl, Niklas Holsti, Stephan
  Thesing, David Whalley, Guillem Bernat, Christian Ferdinand, Reinhold
  Heckmann, Tulika Mitra, Frank Mueller, Isabelle Puaut, Peter Puschner, Jan
  Staschulat, and Per Stenstr\"{o}m.
\newblock The worst-case execution-time problem---overview of methods and
  survey of tools.
\newblock {\em ACM Trans. Embed. Comput. Syst.}, 7(3):36:1--36:53, May 2008.

\end{thebibliography}
\end{document}